\documentclass[twocolumn,showpacs,preprintnumbers,amsmath,amssymb]{revtex4}

\usepackage{graphicx}
\usepackage{dcolumn}
\usepackage{bm}
\usepackage{amsmath}

\begin{document}


\title{Evaporating dynamical horizon with Hawking effect in Vaidya spacetime}

\author{Shintaro Sawayama}
 \email{shintaro@th.phys.titech.ac.jp}
\affiliation{%
Department of Physics, Tokyo Institute of Technology, Oh-Okayama 1-12-1, Megro-ku, Tokyo 152-8550, Japan\\
}%


\date{\today}
\begin{abstract}
We consider how the mass of the black hole decreases by the Hawking radiation in the Vaidya spacetime, 
using the concept of dynamical horizon equation, proposed by Ashtekar and Krishnan.
Using the formula for the change of the dynamical horizon, we derive an equation for the mass incorporating the Hawking radiation.
It is shown that final state is the Minkowski spacetime in our particular model.
\end{abstract}
\pacs{04.25.Dm, 04.70.Bw}
\maketitle
\section{Introduction}
\ \ \ In the study of black hole evaporation, there has been an important issue how black hole mass decreases 
as a back reaction of the Hawking radiation\cite{H}.
We have to confront with this issue to resolve the information loss paradox.
There have been many works concerning black hole evaporation,
either in string theories\cite{HK}\cite{GR}\cite{TT},
or semiclassical theory typically using apparent horizon \cite{PA}.
Hiscok studied spherical model of the black hole evaporation using the Vaidya metric, which we also use in present work, 
to solve the black hole evaporation problem.
However, he simply set a model not taking account of the field equation. 
Hajicek's work\cite{Haji} treated the black hole mass more generally than our present case. 
However, he did not use the field equation either. 
One of the more recent studies is Sorkin and Piran's work \cite{SP} on charged black holes.
And neutral case has been done by Hamade and Stewart\cite{HS}.
Their conclusion is that black hole mass decreases or increases depending on initial condition.
They used a model of the double null coordinates, and obtained a numerical result.
But they did not consider the Hawking effect directly but they used massless scalar field as a matter.
Brevik and Halnes calculated primordial black hole evaporation\cite{BH}.
Very recently Hayward studied black hole evaporation and formation using the Vaidya metric \cite{Hay}.  
It seems no analytical equation has been proposed for the black hole mass with the Hawking effect taken into account.\\
\ \ \ The dynamics for the black hole mass with the Hawking effect is a long standing problem.
Page\cite{Page} derived the equation of mass intuitively, that is $\dot{M}\propto -M^{-2}$.
But it does not come from the first principle. 
We will comment on his intuitive result in the final section. 
To derive the equation of mass from the first principle we should treat the Einstein equation with the back reaction term by the Hawking radiation.
However, the Einstein equation cannot be analytically solved, 
because the equation contains fourth derivative terms as back reaction. 
Recently Ashtekar and Krishnan derived an equation which describes how the horizon changes in time.
It needs only information of the horizon surface.\\
\ \ \  In section \ref{sec2}, dynamical horizon is reviewed. 
In section \ref{sec3}, the location of the dynamical horizon in the Vaidya spacetime is identified. And then in section \ref{sec4}, 
the dynamical horizon equation is written down in the case of Vaidya matter with the Hawking effect being taken into account.
Section \ref{sec5} is devoted to conclusion and discussions.   
\section{Dynamical Horizon} \label{sec2}
\ \ \ Ashtekar and Krishnan considered dynamical horizon \cite{AK1}\cite{AK2},  
and derived a new equation that dictates how the dynamical horizon radius changes.
Apparent horizon is a time slice of the dynamical horizon. 
The definition of dynamical horizon is, \\ \\
{\it Definition}. A smooth, three-dimensional, spacelike submanifold $H$ in a space-time
is said to be a {\it dynamical horizon} if it is foliated by preferred family of 2-spheres such that, 
on each leaf S, the expansion $\Theta _{(l)}$ of a null normal $l^a$ vanishes and the expansion $\Theta_{(n)}$
 of the other null normal $n^a$ is strictly negative. \\ \\
\ \ \ The requirement that one of the null expansions is zero comes from the intuition that black hole does not emit even light.
And the requirement that other null expansion is strictly negative comes from that null matter goes in black holes inwards.\\
\ \ \ In this definition, we can recapitulate the important formula which gives a change of the dynamical horizon radius by the matter flow, 
using 3+1 and then 2+1 decompositions and also the Gauss-Bonet theorem.
\begin{eqnarray}
(\frac{R_2}{2G}-\frac{R_1}{2G})=\int _{\Delta H}T_{ab}\hat{\tau}^a\xi ^b_{(R)}d^3V\nonumber \\
+\frac{1}{16\pi G}\int _{\Delta H}(|\sigma |^2+2|\zeta |^2)d^3V. \label{fdyna}
\end{eqnarray}
Here $R_2,R_1$ the radius of the dynamical horizon
, $T_{ab}$ is the stress-energy tensor, $|\sigma |^2=\sigma_{ab}\sigma^{ab}, |\zeta |^2=\zeta_a\zeta^a$ where $\sigma_{ab}$ is the 
shear, and 
$\zeta ^a:=\tilde{q}^{ab}\hat{r}^c\nabla _c l_b$, with two dimensional metric $\tilde{q}_{ab}$, and 
$\xi ^a_{(R)}:=N_Rl^a \ (N_R:=|\partial R|)$ where $R$ is the radius of the dynamical horizon. \\
\ \ \ This is the dynamical horizon equation that tells how the horizon radius changes by the matter flow, shear and expansion.
In the spherically symmetric case that we shall consider in what follows the second term of the right hand side vanishes. 
Although in the case of quantum field theory in curved space time, the dominant energy condition does not hold\cite{Ford}\cite{BD}, 
we can use the dynamical horizon equation because it is valid even when the black hole radius decreases. 
And the dynamical horizon equation is a consequence of the Einstein equation. 
We use the dynamical horizon equation in place of the Einstein equation. \\
\ \ \ Because of the negative energy, we can expand the definition of the dynamical horizon in the case of timelike. 
Then the definition of the dynamical horizon is now, \\ \\
{\it Definition (modified version)}. A smooth, three-dimensional, spacelike or {\bf timelike} submanifold $H$ in a space-time
is said to be a {\it dynamical horizon} if it is foliated by preferred family of 2-spheres such that, 
on each leaf S, the expansion $\Theta _{(l)}$ of a null normal $l^a$ vanishes and the expansion $\Theta_{(n)}$
 of the other null normal $n^a$ is strictly negative. \\ \\
We can easily calculate the timelike dynamical horizon equation only replacing $\hat{r}^a$ and $\hat{\tau}^a$. 
And the way of 3+1 and 2+1 decomposition is replaced. 
The timelike dynamical horizon was defined by Ashtekar and Galloway and called {\it timelike membrane} \cite{AG}. 
\section{Vaidya Spacetime}\label{sec3}
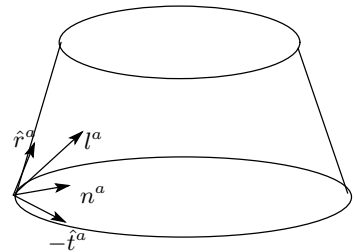
\begin{figure}
\unitlength 0.1in
\begin{picture}( 17.7000, 12.1000)(  7.9000,-18.5000)
%
\special{pn 8}%
\special{ar 1660 830 630 190  0.0000000 6.2831853}%
%
\special{pn 8}%
\special{ar 1680 1640 880 210  0.0000000 6.2831853}%
%
\special{pn 8}%
\special{pa 1040 830}%
\special{pa 800 1640}%
\special{fp}%
%
\special{pn 8}%
\special{pa 2280 860}%
\special{pa 2540 1620}%
\special{fp}%
%
\special{pn 8}%
\special{pa 790 1630}%
\special{pa 900 1360}%
\special{fp}%
\special{sh 1}%
\special{pa 900 1360}%
\special{pa 856 1414}%
\special{pa 880 1410}%
\special{pa 894 1430}%
\special{pa 900 1360}%
\special{fp}%
%
\special{pn 8}%
\special{pa 790 1630}%
\special{pa 1060 1770}%
\special{fp}%
\special{sh 1}%
\special{pa 1060 1770}%
\special{pa 1010 1722}%
\special{pa 1014 1746}%
\special{pa 992 1758}%
\special{pa 1060 1770}%
\special{fp}%
%
\special{pn 8}%
\special{pa 800 1620}%
\special{pa 1150 1300}%
\special{fp}%
\special{sh 1}%
\special{pa 1150 1300}%
\special{pa 1088 1330}%
\special{pa 1112 1336}%
\special{pa 1114 1360}%
\special{pa 1150 1300}%
\special{fp}%
%
\special{pn 8}%
\special{pa 790 1630}%
\special{pa 1080 1580}%
\special{fp}%
\special{sh 1}%
\special{pa 1080 1580}%
\special{pa 1012 1572}%
\special{pa 1028 1590}%
\special{pa 1018 1612}%
\special{pa 1080 1580}%
\special{fp}%
\put(11.6000,-13.9000){\makebox(0,0)[lb]{$l^a$}}%
\put(7.9000,-13.8000){\makebox(0,0)[lb]{$\hat{r}^a$}}%
\put(9.7000,-19.3000){\makebox(0,0)[lb]{$-\hat{t}^a$}}%
\put(11.4000,-16.7000){\makebox(0,0)[lb]{$n^a$}}%
\end{picture}
\caption{For the case that the dynamical horizon decreases, we should choose $l^a=-\hat{t}^a+\hat{r}^a$ 
so that $l^a$ points into the dynamical horizon.}\label{fig5}
\end{figure}

\ \ \ The Vaidya metric is of the form
\begin{eqnarray}
ds^2=-Fdv^2+2Gdvdr+r^2d\Omega ^2 ,
\end{eqnarray}
where $F$ and $G$ are functions of $v$ and $r$,
and $v^a$ is null vector and $r$ is the area radius, and $M$ is the mass defined by $M=\frac{r}{2}(1-\frac{F}{G^2})$, 
a function of $v$ and $r$. 
This metric is spherically symmetric.
By substituting the Vaidya metric (2) into the Einstein equation so that we can identify the energy-momentum tensor $T_{ab}$ as  
\begin{eqnarray}
8\pi T_{vv}&:=&\frac{2}{r^2}(FM_{,r}+GM_{,v}) \\
8\pi T_{vr}&:=&-\frac{2G}{r^2}M_{,r} \\
8\pi T_{rr}&:=&\frac{2G_{,r}}{rG}.
\end{eqnarray}
We do not need to check that the solution of the dynamical horizon equation satisfies the Einstein equation.
Because we would like to consider the Schwarzschild like metric, we set $v=t+r^*$, 
where $r^*$ is tortoise coordinate with dynamics
\begin{eqnarray}
r^*=r+2M(v)\ln \bigg( \frac{r}{2M(v)}-1\bigg) .
\end{eqnarray}
For later convenience, we write,
\begin{eqnarray}
a=\frac{d r}{d r^*}.
\end{eqnarray}
There are two null vectors,
\begin{eqnarray}
l^a=\begin{pmatrix}
l^t \\
l^{r*}\\
l^{\theta} \\
l^{\phi}
\end{pmatrix}=
\begin{pmatrix}
-a^{-1}\\
a^{-1}\\
0\\
0
\end{pmatrix},
\end{eqnarray}
corresponding to the null vector $v^a$,
and the other is
\begin{eqnarray}
n^a=\begin{pmatrix}
n^t\\
n^{r*}\\
n^{\theta}\\
n^{\phi}
\end{pmatrix}=
\begin{pmatrix}
-a^{-1}\\
-\frac{F}{F-2Ga}a^{-1}\\
0\\
0
\end{pmatrix}.
\end{eqnarray}
Here we multiply $a^{-1}$ so that $l^a=v^a$. 
This choice of the null vector $l^a$ is explained in figure \ref{fig5}.
From now on we put,
\begin{eqnarray}
F&=&\bigg( 1-\frac{2M(v)}{r}\bigg) \\
G&=&1,
\end{eqnarray}
in a similar form to the Schwarzschild metric, 
assuming that $M(v)$ is a function of $v$ only. 
For a constant $M$, the metric coincides with the Schwarzschild metric.
We calculate the expansions $\Theta _{(l)}$ and $\Theta _{(n)}$ of the two null vectors $l^a,n^a$,  
because the definition of the dynamical horizon requires one of the null expansions to be zero and the other to be minus. 
The result is,
\begin{eqnarray}
\Theta _{(l)}&=&\frac{1}{r}(2F-a)\\
\Theta _{(n)}&=&\frac{1}{ar}\bigg( \frac{-2F^2+aF-2a^2}{-F+2a}\bigg) .
\end{eqnarray}
From $\Theta _{(l)}=0$ we get,
\begin{eqnarray}
2F-a=0.
\end{eqnarray}
we can check that the other null expansion $\Theta_{(n)} $is strictly negative. 
Therefore in this case, we can apply the dynamical horizon equation. 
In the usual Schwarzschild metric with dynamics, both expansions become zero. 
This is the one of the reasons why we choose the Vaidya metric. 
By inserting equation (7) to equation (14), we obtain
\begin{eqnarray}
a&=&F\bigg( 1-2M_{,v}\ln \bigg(\frac{r}{2M}-1\bigg) \nonumber \\
&+& \frac{r}{M(r/2M-1)}M_{,v}\bigg) .
\end{eqnarray}
Note that $a$ is proportional to $F$. 
Now we solve $\Theta _{(l)}=0$, to determine the dynamical horizon radius as
\begin{eqnarray}
2F-a&=&2F\nonumber \\
&-&F\bigg( 1-2M_{,v}\ln \bigg(\frac{r}{2M}-1\bigg) \nonumber \\
&+& \frac{r}{M(r/2M-1)}M_{,v}\bigg) \nonumber \\
&=&0.
\end{eqnarray}
From this equation we obtain,
\begin{eqnarray}
1+\bigg( -2M_{,v}\ln \bigg(\frac{r_D}{2M}-1\bigg) \nonumber \\
+ \frac{r_D}{M(r_D/2M-1)}M_{,v}\bigg) =0.
\end{eqnarray}
Here $F=0$ is also the solution of the dynamical horizon.
The dynamical horizon radius $r_D$ is given by solving (17) as
\begin{eqnarray}
r_D=2M+2Me^{-v/2M}. \label{f43}
\end{eqnarray}
Note that this dynamical horizon radius is outside the $r=2M$, that is othor solution.
\section{Dynamical Horizon Equation with Hawking Radiation} \label{sec4}
\begin{figure}
\includegraphics[width=50mm]{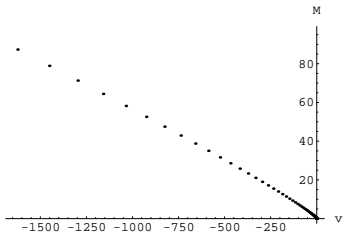}
\caption{Numerical calculation of the black hole mass $M$ as a function of v from the equation (60)}\label{plot}
\end{figure}

\ \ \ At first, we should derive the energy-momentum tensor $T_{\hat{t}l}$ 
for the integration of the dynamical horizon equation. 
For this end we derive it from the given Vaidya matter. 
For $G=1, \ F=1-\frac{2M(v)}{r}$, the non-vanishing components of the energy-momentum tensor becomes 
\begin{eqnarray}
T_{vv}&=&\frac{1}{4\pi r^2}(FM_{,r}+M_{,v}) \label{f45}\\
T_{lr^*}&=&-\frac{1}{4\pi r^2}M_{,r}a \label{f46}\\
T_{r^*r^*}&=&0.
\end{eqnarray}
Here we have made the coordinate transformation from $r$ to $r^*$.
Writing $T_{tl}$ in terms of $T_{vv}$ and $T_{vr^*}$ given by (\ref{f45})(\ref{f46}) with $l^a=v^a$, we see
\begin{eqnarray}
T_{tl}&=&-T_{vv}+T_{vr^*} \nonumber \\
&=&\frac{1}{4\pi r^2}(-FM_{,r}-M_{,v}-aM_{,r}) \nonumber \\
&=&-\frac{1}{4\pi r^2}\frac{5}{2}M_{,v}.
\end{eqnarray}
With $\hat{t}^a$ being the unit vector in the direction of $t^a$, we obtain
\begin{eqnarray}
T_{\hat{t}l}=-\frac{1}{4\pi r^2}\frac{5}{2}M_{,v}F^{-1}.
\end{eqnarray}
For the dynamical horizon integration (\ref{fdyna}), 
we get
\begin{eqnarray}
\int _{r_1}^{r_2}4\pi r_D^2T_{\hat{t}l}dr_D=\frac{5}{2}\int _{M_1}^{M_2}(1+e^{-v/2M})dM, \label{f52}
\end{eqnarray}
where we have used
\begin{eqnarray}
F=\frac{e^{-v/2M}}{1+e^{-v/2M}},
\end{eqnarray}
and the fact
\begin{eqnarray}
\frac{dM}{dv}=-e^{-v/2M}\bigg( 2(1+e^{-v/2M})+\frac{v}{M}e^{-v/2M}\bigg) ^{-1},
\end{eqnarray}
changing the integration variable from $r_D$ to $M$. 
In the above calculation, we treat $M_{,v}$ and $F^{-1}$ with $r_D$ fixed, 
because these functions are used only in the integration.
Inserting equation (\ref{f52}) to the dynamical horizon equation (\ref{fdyna}), we obtain
\begin{eqnarray}
\frac{1}{2}(2M+2Me^{-v/2M})\bigg| _{M_1}^{M_2} \nonumber \\
=\int _{M_1}^{M_2}\frac{5}{2}(1+e^{-v/2M})dM .
\end{eqnarray}
Taking the limit $M_2\to M_1=M$, we obtain
\begin{eqnarray}
-\frac{3}{2}(1+e^{-v/2M})+\frac{v}{2M}e^{-v/2M}=0 .
\end{eqnarray}
This equation is the dynamical horizon equation in the case that only the Vaidya matter is present. 
There is no solution of this equation, except the trivial one ($F=0$ or $r=2M$), so
\begin{eqnarray}
r_D=2M(v).
\end{eqnarray}
Here $M(v)$ is the arbitrary function only of the $v$, which represent the Vaidya black hole spacetime. \\
\ \ \ Next, we take into account the Hawking radiation. 
To solve this problem, we use two ideas that is to use the dynamical horizon equation, and 
to use the Vaidya metric. 
The reason to use the dynamical horizon equation comes from the fact that we need only information of matter near horizon,  
without solving the full Einstein equation with back reaction being the fourth order differential equations, 
for a massless scalar field.
For the matter on the dynamical horizon, we use the result of Candelas \cite{C}, 
which assumes that 
spacetime is almost static and is valid near the horizon, $r\sim 2M$.
\begin{eqnarray}
T_{tl}&=&-T_{tt}\nonumber \\
&=&\frac{1}{2\pi ^2(1-2M/r)}\int _0^{\infty}\frac{d\omega \omega ^3}{e^{8\pi M\omega}-1} \nonumber \\
&=&\frac{1}{2cM^4\pi ^2(1-2M/r)},
\end{eqnarray}
where we have used a well known result, 
\begin{eqnarray}
\int _0^{\infty}\frac{d\omega \omega ^3}{e^{a\omega}-1}=\frac{\pi ^4}{15a^4},
\end{eqnarray}
and where $c=61440$.
This matter energy is negative near the event horizon. 
In the dynamical horizon equation, if black hole absorbs negative energy, 
black hole radius decreases. 
This is one of the motivations to use the negative energy tensor. 
Next we replace length of $t$ to unit length, because in the dynamical horizon equation $\hat{t}$ is used, so
\begin{eqnarray}
\hat{t}^{0}=F^{-1/2},\ \ l^{0}=F^{-1/2},
\end{eqnarray}
and therefore, the energy tensor becomes
\begin{eqnarray}
T_{\hat{t}l}=\frac{1}{2M^4c\pi ^2(1-2M/r)^2}.
\end{eqnarray}
Calculating the integration on the right hand side of (\ref{fdyna}) for this matter,
\begin{eqnarray}
b\int \frac{r_D^2}{M^4(1-2M/r_D)^2}dr_D\nonumber \\
=b\int \frac{4M^2(1+e^{-v/2M})^4}{M^4e^{-v/M}}\frac{dr_D}{dM}dM \nonumber \\
=b\int _{R_1}^{R_2}\frac{4(1+e^{-v/2M})^4}{M^2}e^{-v/M}\nonumber \\
\times \bigg( 2(1+e^{-v/2M})+\frac{v}{M}e^{-v/2M}\bigg) dM.
\end{eqnarray}
Here we insert the expression for $r_D$ (\ref{f43})in the first line, and the expression for $dr_D/dM=2(1+e^{-v/2M})+\frac{v}{M}e^{-v/2M}$ is used. 
Here $b$ is a constant calculated in \cite{C}
\begin{eqnarray}
b=\frac{1}{30720\pi}.
\end{eqnarray}
If we also take account of the contribution of the Vaidya matter, 
and inserting this into the integration to the dynamical horizon equation (\ref{fdyna}), we obtain
\begin{eqnarray}
\frac{1}{2}( 2M+2Me^{-v/2M})\bigg| _{M_1}^{M_2}\nonumber \\
=b\int _{M_1}^{M_2}\frac{2^2(1+e^{-v/2M})^4}{M^2}e^{-v/M}\nonumber \\
\times \bigg( 2(1+e^{-v/2M})+\frac{v}{M}e^{-v/2M}\bigg) dM \nonumber \\
+\int _{M_1}^{M_2}\frac{5}{2}(1+e^{-v/2M})dM.
\end{eqnarray}
Taking the limit $M_2\to M_1=M$, we finally get
\begin{eqnarray}
-\frac{3}{2}(1+e^{-v/2M})+\frac{v}{2M}e^{-v/2M}\nonumber \\ 
=b\frac{2^2(1+e^{-v/2M})^4}{M^2}\nonumber \\
\times e^{v/M}\bigg( 2(1+e^{-v/2M})+\frac{v}{M}e^{-v/2M}\bigg) , 
\end{eqnarray}
or
\begin{eqnarray}
M^2=\frac{8b(1+e^{-v/2M})^4e^{v/M}}{-\frac{3}{2}(1+e^{-v/2M})+\frac{v}{2M}e^{-v/2M}}\nonumber \\
\times \bigg( (1+e^{-v/2M})+\frac{v}{2M}e^{-v/2M}\bigg) .\label{f4}
\end{eqnarray}
This is the main result of the present work that describes 
how the mass of black hole decreases. 
This equation is the transcendental equation, so usually it cannot be solved analytically. 
However, with the right hand side depending only on $-v/2M$, we can easily treat Eq.(\ref{f4}) numerically. 
Figure \ref{plot} is a graph of $M$ as a function of $v$ \\
\ \ \ If the dynamical horizon were inside the event horizon, the dynamical horizon radius would be
\begin{eqnarray}
r_D=2M-2Me^{-v/2M}.\nonumber
\end{eqnarray}
In this case, the dynamical horizon equation would become
\begin{eqnarray}
M^2=\frac{-8b(1-e^{-v/2M})^4e^{v/M}}{\frac{7}{2}(1-e^{-v/2M})-\frac{v}{2M}e^{-v/2M}}\nonumber \\
\times \bigg( (1-e^{-v/2M})-\frac{v}{2M}e^{-v/2M}\bigg) .\nonumber
\end{eqnarray}
The singular behavior of this expression excludes its physical relevance. \\
\ \ \ Now we show an approximation of the Eq.(\ref{f4}) in particular limiting case.
Taking the limit $M\to 0$, and $-v/2M={\rm const}$, we can see that (\ref{f4}) becomes,
\begin{eqnarray}
\frac{bC_1}{M^2}+\frac{bvC_2}{M^3}=0.
\end{eqnarray}
Where $C_1,C_2$ are positive constants. or
\begin{eqnarray}
M=-\frac{C_2v}{C_1}.
\end{eqnarray}
So, in the vanishing process the mass is proportional to $v$.
For $M \to {\rm large}$ 
\begin{eqnarray}
\dot{M}=-\frac{C_3}{\log M},
\end{eqnarray}
where $C_3$ is a positive constant.
It comes from the limit $M\to \infty$ and $-v/2M\to \infty$. 
In this limit the equation (\ref{f4}) becomes $v=-2M\log M$. 
This is different from Page's result \cite{Page}. 
Because if $M$ goes to large, the dynamical horizon radius increases as $M^{2}$, 
so absorbed energy also become large. 
From this reason derivative of $M$ by $v$ changes. 
If we do not consider next order, the derivative of $M$ becomes $\dot{M}=-C_4$,
so that $4\pi r_D^2T^4\approx 1$, contradicting with Page's intuition.  \\
\section{Conclusion and Discussions}\label{sec5}
\ \ \ We have derived an equation which describes how the black hole 
mass changes taking into account of the Hawking radiation, in the special 
Vaidya spacetime which becomes the Schwarzschild spacetime in the static case. 
From the analysis of the transcendental equation (38), 
we have shown that the black hole mass eventually vanishes and the spacetime becomes the Minkowski spacetime independent of 
the initial black hole mass size.\\
\ \ \ The dynamical horizon method in this paper can take into account of 
the back reaction of the Hawking radiation without solving the field equation which contains the fourth order differentials. \\
\ \ \  In the limit of the black hole mass going to zero, the derivative of the mass becomes small in 
proportion to the null coordinate ($v=t+r^*$). 
On the other hand as the black hole mass becomes large, the derivative behaves the minus of the 
inverse of the logarithm of the mass. Our result, which is different from Page's result, 
comes from the fact that in the large mass limit, 
the black hole radius behaves like quadratic of the black hole mass. 
This probably comes from when large mass limit that the approximation $r\to 2M$ is broken.\\
\ \ \ We would like to compare the present work to the preceding works.  
Sorkin and Piran or Hamade and Stewart used a massless 
scalar field instead of the Hawking radiation as the back reaction directly. 
The conclusion of their paper is that black hole starts with the small mass 
and it evaporates or increases. 
However, it is shown in the present work that even if the black hole starts with a 
large mass it always vanishes. \\
\ \ \ Although we have treated the black hole evaporation semi-classically, 
we hope this work will give an intuition to quantization of black holes. 
\begin{acknowledgments}
We would like to acknowledge A.Hosoya, T.Mishima, T.Okamura, and M.Shiino for comments and discussions. 
\end{acknowledgments}


\begin{thebibliography}{99}
\bibitem{H}
S.W.Hawking {\em Commun. Math. Phys.} {\bf 43} 199 (1975)
\bibitem{HK}
C.M.Harris and P.Kanti {\em JHEP} 0310 014 hep-ph/0309054 (2003)
\bibitem{GR}
J.G.Russo hep-th/0501132
\bibitem{TT}
T.Tanaka {\em Prog.Theor.Phys.Suppl} {\bf 148} 307 (2003)
\bibitem{PA}
P.Anninos et. al. {\em Phys. Rev. Lett.} {\bf 74} 630 (1995)
\bibitem{Hisc}
W.A.Hiscock {\em Phys. Rev.} {\bf D 23} 2813 (1981)
\bibitem{Haji}
P.Hajicek {\em Phys. Rev.} {\bf D 36} 1065 (1987)
\bibitem{SP}
E.Sorkin and T.Piran {\em Phys. Rev.} {\bf D 63} 124024 (2001)
\bibitem{HS}
R.S.Hamade and J.M.Stewart {\em Class. Quantum Grav.} {\bf 13} 497 (1996)
\bibitem{BH}
I.Brevik and G.Halnes {\em Phys. Rev} {\bf D 67} 023508 (2003)
\bibitem{Hay}
S.A.Hayward gr-qc/0506126
\bibitem{Page}
D.N.Page {\em Phys. Rev.} {\bf D 13} 198 (1976)
\bibitem{AK1}
A.Ashtekar, B.Krishnan {\em Phys. Rev. Lett} {\bf 89} 261101 (2002)
\bibitem{AK2}
A.Ashtekar, B.Krishnan gr-qc/0407042 (2005)
\bibitem{AG}
A.Ashtekar, G.J.Galloway gr-qc/0503109 (2005)
\bibitem{Ford}
L.H.Ford gr-qc/9707062 (1997)
\bibitem{BD}
N.D.Birrell and P.C.W.Davies {\em Quantum fields in curved space} Cambridge University Press (1982)
\bibitem{C}
P.Candelas {\em Phys. Rev.} {\bf D 21} 2185 (1980)
\end{thebibliography}
\end{document}